\documentclass[reprint, superscriptaddress, amsmath,amssymb,
 aps]{revtex4-2}
\usepackage{graphicx}
\usepackage{bbm}
\usepackage{physics}
\usepackage{amssymb}
\usepackage{amsmath}
\usepackage{dcolumn}
\usepackage{bm}
\usepackage{lipsum}
\usepackage{braket}
\usepackage{siunitx}

\usepackage{epigraph} 
\usepackage{hyperref}

\makeatletter
\@ifundefined{pdfinclusioncopyfonts}{}{%
  \pdfinclusioncopyfonts=1
}
\makeatother

\usepackage{xcolor}
\graphicspath{{figs/}}

\makeatletter
\@ifundefined{switch@array}{}{%
  \def\switch@array{}%
}
\makeatother

\begin{document}

\preprint{APS/123-QED}
\title{
Demonstrating and Benchmarking Classical Shadows for Lindblad Tomography
}

\author{Rune Thinggaard Birke}
\email{ruth@math.ku.dk}
\affiliation{Center for Quantum Devices, Niels Bohr Institute, University of Copenhagen, Denmark}
\affiliation{Center for the Mathematics of Quantum Theory, MATH Department, University of Copenhagen}
\affiliation{NNF Quantum Computing Programme, Niels Bohr Institute, University of Copenhagen, Denmark}

\author{Johann Bock Severin}
\affiliation{Center for Quantum Devices, Niels Bohr Institute, University of Copenhagen, Denmark}
\affiliation{NNF Quantum Computing Programme, Niels Bohr Institute, University of Copenhagen, Denmark}

\author{Malthe A. Marciniak}
\affiliation{Center for Quantum Devices, Niels Bohr Institute, University of Copenhagen, Denmark}
\affiliation{NNF Quantum Computing Programme, Niels Bohr Institute, University of Copenhagen, Denmark}

\author{Emil Hogedal}
\author{Andreas Nylander}
\author{Irshad Ahmad}
\author{Amr Osman}
\author{Janka Biznárová}
\author{Marcus Rommel}
\author{Anita Fadavi Roudsari}
\author{Jonas Bylander}
\author{Giovanna Tancredi}
\affiliation{Department of Microtechnology and Nanoscience, Chalmers University of Technology, SE-412 96 Gothenburg, Sweden}

\author{Daniel Stilck França}
\affiliation{Center for the Mathematics of Quantum Theory, MATH Department, University of Copenhagen}
\author{Albert Werner}
\affiliation{Center for the Mathematics of Quantum Theory, MATH Department, University of Copenhagen}
\author{Christopher W. Warren}
\affiliation{Center for Quantum Devices, Niels Bohr Institute, University of Copenhagen, Denmark}
\affiliation{NNF Quantum Computing Programme, Niels Bohr Institute, University of Copenhagen, Denmark}

\author{Jacob Hastrup}
\affiliation{Center for Quantum Devices, Niels Bohr Institute, University of Copenhagen, Denmark}
\affiliation{NNF Quantum Computing Programme, Niels Bohr Institute, University of Copenhagen, Denmark}

\author{Svend Krøjer}
\affiliation{Center for Quantum Devices, Niels Bohr Institute, University of Copenhagen, Denmark}
\affiliation{NNF Quantum Computing Programme, Niels Bohr Institute, University of Copenhagen, Denmark}

\author{Morten Kjaergaard}
\email{mkjaergaard@nbi.ku.dk}
\affiliation{Center for Quantum Devices, Niels Bohr Institute, University of Copenhagen, Denmark}
\affiliation{NNF Quantum Computing Programme, Niels Bohr Institute, University of Copenhagen, Denmark}

\date{\today}

\begin{abstract}
Spurious couplings and decoherence degrade the performance of solid‑state quantum processors, demanding careful design, calibration, and mitigation protocols. These strategies often rely on characterization of the idling processor, but tomographic recovery of (time‑independent) Lindblad dynamics scales exponentially with qubit count. Here, we experimentally benchmark and demonstrate that randomized (“shadow”) measurements accelerate Lindblad tomography on a superconducting transmon processor. We first implement \textit{extensible Lindblad tomography}, which estimates Lindblad parameters using a complete tomographic dataset, and use it as a baseline to benchmark a shadow tomography approach, \textit{shadow Lindblad tomography}. The shadow approach recycles randomized configurations to estimate the same Lindblad parameters using far fewer resources under physically motivated locality assumptions. We experimentally verify these assumptions in our processor by implementing the protocols on one‑ and three‑qubit subsystems; here, shadow Lindblad tomography reproduces extensible Lindblad tomography within uncertainties while using substantially fewer configurations. Leveraging this efficiency, we apply shadow Lindblad tomography to the full five‑qubit processor and recover all single qubit dissipation and two-qubit coupling parameters in 9 hours of acquisition time compared to an estimated 58 hours for extensible Lindblad tomography.
Additionally, our shadow implementation is compatible with conventional Gaussian error propagation, avoiding the use of median-of-means estimators. Together, these results demonstrate how randomized shadow tomography protocols can be practically implemented to learn quantum processor dynamics at an increasing qubit count.
\end{abstract}

\maketitle


\section{Introduction}
Precise calibration of quantum processors requires the reliable estimation of dynamical parameters under realistic device noise \cite{eisert_quantum_2020}. 
Beyond calibration, characterization informs processor design and fabrication, enabling the development of higher-performance devices \cite{kjaergaard_superconducting_2020}. 
Yet, exhaustive characterization techniques, such as tomographic techniques, scale exponentially with system size, quickly overwhelming experimental resources \cite{cotler_quantum_2020, cramer_efficient_2010, gross_quantum_2010, binosi_tailor-made_2024, titchener_scalable_2018}.
This necessitates physics-motivated assumptions about the dynamics of quantum systems, ensuring that characterization protocols remain experimentally tractable.

One of the most prominent leaps within characterization and quantum tomography has been the development of shadow tomography \cite{aaronson_shadow_2018, Huang_shadows_2020}. 
Shadow tomography is a term coined for tomographical procedures that attempt to estimate many expectation values via random states or measurement bases \cite{hadfield_adaptive_2021, acharya_shadow_2021, struchalin_experimental_2021, zhao_fermionic_2021, chen_robust_2021, garcia_quantum_2021}. 
Under the right choice of random configurations, shadow tomography procedures are found to have exponential speed-ups for extracting information compared to their deterministic counterparts. Here, relevant examples count low-weight expectation values of states \cite{Huang_shadows_2020} and channels \cite{levy_shadowQPT_2024} or non-local properties such as fidelity \cite{Randomized_measurement_toolbox}.
Experimental implementations of classical‑shadow protocols have been realized and benchmarked on multiple platforms—including photonic systems \cite{Struchalin2021PRXQuantum} and trapped‑ion processors \cite{elben_cross_platform_2020, Seif2023PRXQuantum, ShadowQST}, with observed performance consistent with theoretical predictions. The cost of this proposed speed-up, however, is that one either imposes assumptions about the physical system in question or restricts the class of learnable parameters. 

One way to satisfy the necessary conditions for shadow tomography is to assume physically motivated minimal models~\cite{Franca2024Efficient}. Such physics-informed models are especially relevant for superconducting processors, where long-range and many-qubit effects are typically taken to be negligible~\cite{Acharya2023}. Thus, extracting the idling dynamics of a sparse Lindblad operator can be performed relatively inexpensively in a process tomography setting~\cite{chuang_prescription_1997, nielsen_and_chuang, samach2022lindblad,Franca2025_time_dependent}. Such protocols, referred to as Lindblad tomography, capture open‑system effects by directly learning the idling time-evolution, i.e., the Hamiltonian and dissipators, under a Markovian model~\cite{samach2022lindblad, berg_large_scale_2025}. 
In superconducting devices, it has been experimentally shown to recover coherent couplings and quantify crosstalk. \cite{samach2022lindblad, TensorEsprit_2024, berg_large_scale_2025}
This contrasts with gate‑level SPAM‑robust schemes such as gate‑set tomography (GST), which learn a self‑consistent description of a chosen gate at the cost of being exponentially expensive~\cite{nielsen_gate_2021,greenbaum_introduction_2015}.
Lindblad tomography offers a practical tool for device‑level metrology as it yields interpretable parameters tied to fabrication and layout, supports performance diagnostics, and supplies noise models for control and verification. 

In this work, we experimentally deploy both non-shadow and shadow-based Lindblad tomography on a superconducting 5-qubit processor. Both variants, termed extensible and shadow Lindblad tomography (ELT/SLT) respectively, are based on the recent proposal in Ref.~\cite{Franca2024Efficient}. The protocols acquire tomographic data at short times to recover the Lindblad parameters describing the idling dynamics. By implementing both protocols on single- and 3-qubit subsystems, we reconstruct all Lindblad parameters, including vanishing three-body interactions and correlated decay, experimentally verifying the locality assumptions necessary for SLT to exponentially reduce the resource overhead for larger system sizes. We then deploy SLT on our full 5-qubit processor where ELT is prohibitively expensive even under locality assumptions. In this setting, SLT extracts all single qubit dissipators and two-qubit interactions, paving the way for practical implementations at larger system sizes with only logarithmic overhead.

The article is organized as follows. Section II introduces the general framework of Lindblad tomography. Sections IIa and IIb present ELT and SLT, discussing both their conceptual foundations and practical implementation details. 
Sections III and IV demonstrate the application of these methods to single- and three-qubit subsystems, respectively. 
Section V explores SLT with a restricted model applied to a 5-qubit system. 
Finally, Section VI and VII conclude with discussion and outlook.

\section{Lindblad Tomography}\label{sec:LindbladTomography}
\begin{figure}[t!]
    \centering
        \includegraphics{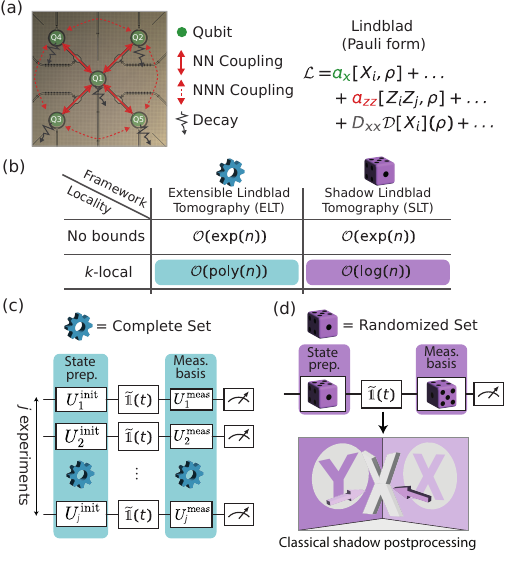}
    \caption{ELT and SLT blueprint. \textbf{(a)} Optical image of a lithographically identical 5-qubit processor to the one used in this work with schematics of coupling and loss mechanisms (left panel).
    A general qubit processor can be described by a Markovian LME with a Lindblad operator, here written out in the Pauli basis (right panel).
    \textbf{(b)} The LME can be learned via ELT or SLT. 
    Both procedures are, in the worst case, exponentially expensive in system size but become feasible under locality assumptions of the underlying LME. 
    \textbf{(c)} ELT extracts Lindblad parameters using a data set that loops over a tomographically complete set of $j$ input and output states and letting the system evolve for varying lengths of time $t$.  
    \textbf{(d)} SLT comprises of initializing a random input state and measurement basis and letting the system evolve for varying lengths of time $t$ and then performing shadow tomographical postprocessing (see Sec.~\ref{sec:SLT_theory} for details) to extract Lindblad parameters.}
    \label{fig: Figure1}
\end{figure}

This section provides the technical details of ELT and SLT, reviewing Ref.~\cite{Franca2024Efficient} in an experimental context.

We aim to learn a time‑independent Lindblad generator consisting of a Hamiltonian $H$ and dissipators $\{L_k\}$ on a 5-qubit superconducting processor, shown schematically in Fig.~\ref{fig: Figure1}a. 
Details of the experimental setup are presented in Appendix \ref{app:setup}.
ELT is a deterministic scheme consisting of a complete tomographic set of product‑state preparations and Pauli measurements (see Fig.~\ref{fig: Figure1}c).
SLT is a shadow tomography scheme that uses randomized state-preparations and measurements along with classical‑shadow post‑processing (see Fig.~\ref{fig: Figure1}d). 
Both procedures scale exponentially in the absence of structure, under physically motivated $k$‑locality (on‑site dissipation and $k$‑local couplings) ELT scales polynomially with system size, but SLT achieves a logarithmic scaling by recycling randomized data across many observables (Fig.~\ref{fig: Figure1}b). 

In the following, we assume a time-independent Lindblad master equation (LME) acting solely on a qubit subspace and access to short-time dynamics, product-state preparations, and Pauli measurements.
The LME governs the time evolution of a state $\rho$ by 
\begin{align}\label{eq:LME}
    \Dot{\rho} &=\mathcal{L}(\rho) = -\frac{i}{\hbar}[H,\rho] + \sum_k L_k\rho L_k^\dagger - \frac{1}{2}\{L_k^\dagger L_k, \rho\},
\end{align}
where $H$ is the system Hamiltonian and $\{L_k\}$ is a set of jump operators that describe dissipation of the system. 
We denote the superoperator $\mathcal{L}$ as the Lindblad of our system. 
To set the stage for Lindblad tomography, we expand our Hamiltonian and jump operators into an $n$-qubit Pauli basis $\mathcal{P}_n \equiv \{I,X,Y,Z\}^{\otimes n}$ as
\begin{equation}
    H = \sum_{P\in \mathcal{P}_n} a_P\, P,\qquad
    L_k = \sum_{Q\in \mathcal{P}_n} c_{k,Q}\, Q.
\end{equation}
Inserting these into Eq.~\eqref{eq:LME} we obtain
\begin{align}\label{eq:LME_v2}
    \mathcal{L}= -i\sum_{P\in\mathcal{P}}a_P[P,\rho] + \sum_{P, Q \in \mathcal{P}}D_{P,Q}\left(P\rho Q - \frac{1}{2}\{QP,\rho\}\right),
\end{align}
where $D_{P,Q} = \sum_{k}c_{k,Q}c_{k,P}^*$ is the dissipation matrix. 

In realistic devices, the Hamiltonian $H$ is dominated by 1‑local (on‑site) and 2‑local (pairwise) terms, but may also contain higher‑order $k$‑local interactions (e.g., three‑body). Likewise, the dissipation matrix $D$ collects decay channels that are typically 1‑local, yet can include correlated multi‑qubit processes (2‑local or higher). A physical representation of these processes can be seen in Fig.~\ref{fig: Figure1}(a). The inclusion of higher-order interactions significantly increases the complexity of the Lindblad and the number of free parameters. To quantify this, note that the Hamiltonian expansion involves $4^n-1$ non-identity Pauli strings, corresponding to $4^n-1$ real parameters. For dissipation, the Hermiticity of $D$ and the tracelessness of the jump operators constrain the number of parameters, but in general, $D$ spans $(4^n-1)^2$ real free parameters. 

In order to collect the parameters of interest, we recast Eq.~\eqref{eq:LME_v2} to gather these coefficients in a vector $\Vec{p}$ and explicitly write the LME as a linear combination of fixed superoperators,
\begin{align}
    \Dot{\rho} = \Vec{p}\cdot \Vec{\mathcal L}(\rho) = \sum_{j}p_j \mathcal{L}_j(\rho).
\end{align}
Each $\mathcal{L}_j$ is either a coherent term $\mathcal{L}_j = -i[P_j,\cdot]$ or a dissipator term of the form $\mathcal{L}_j = P_j(\cdot)Q_j - \tfrac{1}{2}\{Q_jP_j, \cdot\}$. In this representation, learning the dynamics reduces to estimating the coefficients of $\vec{p}$. This makes $\vec{p}$ a vector of $4^n-1$ coherent parameters and $(4^n-1)^2$ incoherent parameters for a total of $4^n(4^n-1)$ Lindblad parameters.

We continue by examining the Lindblad via instantaneous rates of expectation values at $t=0$. To define the signal of interest, consider a collection of input observables $\{A_i\}_{i\in[j]}$ and a collection of output observables $\{B_i\}_{i\in[j]}$, both of size $j$ (typically product states and Pauli operators, but could in general be any Hermitian operators). 
Define
\begin{equation}
    E_i(t) \equiv  \Tr(B_i A_i(t)) = \Tr(B_i\widetilde{\mathbbm{1}}(t)\left(A_i(0)\right)),
\end{equation}
where $\widetilde{\mathbbm{1}}(t)$ denotes idling time-evolution according to the LME for time $t$.  Differentiating at $t=0$ gives 
\begin{align}
    \frac{d}{dt}E_i(0) &= \Tr(B_i\Dot{A}_i(0)) = \Tr(B_i\mathcal{L}(A_i))\big{|}_{t = 0} \\
    &= \sum_j p_j\Tr(B_i\mathcal{L}_j(A_i))\big{|}_{t = 0}.\label{eq:LME_probes}
\end{align}
This can be seen as a system of equations with unknown parameters $p_j$ but known observables $\{A_i\}$ and $\{B_i\}$. 
Collecting all configurations into a vector yields the compact linear relation
\begin{equation}
    \frac{d}{dt}\Vec{E}(0) = \textsf{M}\Vec{p}\label{eq:ELT_eq},
\end{equation}
where $\textsf{M}_{ij} = \Tr(A_i\mathcal{L}_j(B_i))$ is a complex-valued transfer matrix, depending only on the choice of input and output observables. In order to learn the parameters in $\vec{p}$, we can invert the expectation values according to Eq.~\eqref{eq:ELT_eq} writing it out as
\begin{equation}\label{eq:Transformed_PTM_signal}
    P(t) = \textsf{N}\Vec{E}(t) 
\end{equation}
for any choice of matrix $\textsf{N}$ for which $\textsf{N}\textsf{M} = \mathbbm{1}$. We will refer to the time-series $P(t)$ as a transformed Transfer Matrix (transformed TM). We then estimate Lindblad parameters from the rate of the transformed TM at intercept,
\begin{equation}\label{eq:Transformed_PTM_t=0}
    \Vec{p} = \frac{d}{dt}P(t) \big|_{t=0}.
\end{equation}
 
Hence, learnability of parameters requires that $\textsf{M}$ has full column rank over the parameter subset of interest. This translates to a necessary and sufficient condition that can be examined numerically for which configurations should be used to learn all parameters. The condition number of $\textsf{M}$ furthermore controls the statistical stability of the estimated parameters.

Importantly, the sum in Eq.~\eqref{eq:LME_probes} is sparse given the right choice of configurations under physically motivated assumptions on the Lindblad~\cite{Franca2024Efficient}.
Therefore, $E_i(t)$ can be experimentally estimated using short time traces and can be efficiently transformed into a transformed TM according to Eq.~\eqref{eq:Transformed_PTM_signal}. 
These transformed TMs will then be the core observables for parameter recovery; one determines Lindblad parameters, $\vec{p}$, by extracting their slopes at the intercept using low-order polynomial fits. 

The difference between ELT and SLT then consists of which set of operators $\{A_i\}, \{B_i\}$ we use as input to $E_i(t)$ and in how to procure the expectation values $E_i(t)$ in an actual experimental protocol. 

In Appendix~\ref{app:fitting}, we directly verify on a 2-qubit dataset that $\chi^2$-fitting yield statistically consistent parameter estimates with previous robust fitting procedures~\cite{Franca2024Efficient} and use this method for all data analysis.

\begin{figure}[t!]
    \centering
    \includegraphics{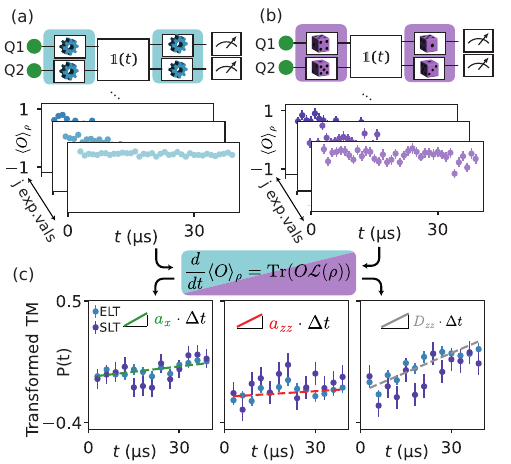}
    \caption{Extracting Lindblad parameters using ELT and SLT for two qubits. 
    \textbf{(a)} Pauli expectation values as a function of time. 
    Each of the $j$ expectation values is found from a single choice of initialization and finalization gates. 
    \textbf{(b)} Transfer matrix entries as a function of time. 
    Each expectation value is found from all randomization experiments (details in Sec.~\ref{sec:SLT_theory}). 
    \textbf{(c)} The tomographic data is linearly transformed and fitted to a low-order polynomial (here a linear fit, see further details in Sec.~\ref{sec:LindbladTomography}). The Lindblad parameters are then extracted as the derivative of the fitted curve at $t=0$.}
    \label{fig: Figure2}
\end{figure}

\subsection{Extensible Lindblad Tomography}\label{sec:ELT_theory}
ELT deterministically implements the Lindblad tomography protocol from above by considering the particularly simple family of configurations that are product eigenstates of Pauli operators for preparation and Pauli measurements for readout. In the language of Section \ref{sec:LindbladTomography}, this means that $\{B_i\}$ are Pauli operators and $\{A_i\}$ are density matrices of pure product eigenstates of Pauli operators. As $\{A_i\}$ in this case is a set of physically implementable states and $\{B_i\}$ is a set of measurable observables, we estimate $E_i(t)$ by simply preparing the states $\{A_i\}$, evolving them for time $t$ and measuring the observables $\{B_i\}$. 

Such Pauli-based constructs are standard in process tomography and minimize experimental complexity of implementation \cite{nielsen_and_chuang}. For this family of configurations, $\textsf{M}$ can be inverted analytically when assuming that the dissipation terms of the master equation only act independently on single qubit sites, neglecting correlated qubit decay. Furthermore each row of $\textsf{N} = \textsf{M}^{-1}$ only contains a few non-zero entries, implying that each Lindblad parameters can be learned from just a few experiments~\cite{Franca2024Efficient}. 

Figure~\ref{fig: Figure1}(c) sketches the necessary acquisition loop of ELT and Fig.~\ref{fig: Figure2}(a,c) illustrate the short-time traces extracted with corresponding slope extraction using qubits 1 and 2 on our 5-qubit processor.
Experimentally, the protocol consists of three steps:
\begin{enumerate}
    \item \textbf{Select configurations:} Choose an informationally complete set of input states $\{\rho_i\}$ (here product Pauli eigenstates) and observables $\{O_i\}$ (here Pauli operators), for a total of $N_\text{conf}$ configurations. Precompute $\textsf{M}_{ij} = \Tr(O_i\mathcal{L}_j(\rho_i)) $ and find any $\textsf{N}$ with $\textsf{N}\textsf{M}=\mathbbm{1}$ for the chosen parameterization.
    \item \textbf{Acquire short-time traces:} For each configuration $i$, initialize $\rho_i$ by implementing an initialization pulse $U_i^{\text{init}}$ and evolve for a set of $N_t$ times $\{t\}$ under the idling dynamics $\widetilde{\mathbbm{1}}(t)$. Finally, measure $O_i$ by applying a finalization gate $U_i^{\text{meas}}$, measure in the computational basis and receive a single-shot estimate of $O_i$. 
    Repeat this $N_\text{shot}$ times and compute the mean of the single-shot estimates to learn the expectation value $E_i(t)$ to target accuracy for each time.
    \item \textbf{Transform and estimate rates by low-polynomial fits:} Transform the data according to Eq.~\eqref{eq:Transformed_PTM_signal}. Perform low-polynomial fits over the time axis to extract Lindblad parameters $\Vec{p}$ from the first derivative at $t=0$. In our experiments, we are in the linear-response regime and thus solely utilize linear fits. 
\end{enumerate} 
This procedure requires a total of $N_\text{conf}N_\text{shot}N_t=N_\text{ELT}N_t$ experimental runs, where $N_\text{ELT}=N_\text{conf}N_\text{shot}$ is the number of runs per time step. Since coherent parameters are real and incoherent parameters are complex, we report the magnitude of extracted parameters in this work for comparability. The uncertainty intervals for these magnitudes are then estimated as $16\%$ and $84\%$ quantiles (corresponding 1-sigma error bars of a normal distribution) of the Rice distribution associated with the errors found from the low-polynomial fit~\cite{Rice_1944}.

This procedure is in general completely model agnostic in the sense that it can be used to learn any parameter of a general Lindblad model. The number of parameters to be estimated determines the number of configurations, leading to exponential scaling with system size in the unconstrained case.
Typically, physically motivated restrictions put some of the possible parameters to zero, e.g. dissipation is assumed to be only on-site.
Such assumptions reduces the problem to having only $\text{poly}(n)$ learnable parameters each needing $O(1)$ time traces to learn, making the protocol provably efficient, see Fig.~\ref{fig: Figure1}(b). 

\subsection{Shadow Lindblad Tomography}\label{sec:SLT_theory}
Shadow tomography refers to a family of randomized protocols that estimate many observables from few experiments by leveraging stochastic input states and measurement bases~\cite{Randomized_measurement_toolbox}.
Here, we explain how we experimentally implement shadow accelerated Lindblad tomography~\cite{Franca2024Efficient}.
As described in section \ref{sec:LindbladTomography}, for Lindblad tomography, one needs to learn a set of expectation values $E_i(t) = \Tr(B_i\widetilde{\mathbbm{1}}(t)(A_i))$, transform them according to Eq.~\eqref{eq:Transformed_PTM_signal} and fit them, in our case, with a linear fit to extract a slope. 
The difference between SLT and ELT is in the input and output observables ($\{A_i\}, \{B_i\}$) used to define $E_i(t)$, and the fact that we acquire the expectation values in ELT via deterministic choice of state preparation and measurement bases, whereas SLT uses classical shadows for process tomography at each evolution time to simultaneously estimate many expectation values by recycling data from randomized configurations. 
In SLT, the input and output observables of interest are Pauli operators, meaning that $\{A_i\}, \{B_i\}$ are all Pauli strings,
\begin{equation}
    E_{P,Q}(t) \equiv \tfrac{1}{2^n}\,\Tr(P\widetilde{\mathbbm{1}}(t)(Q)),\label{eq:PTM_SLT}
\end{equation}
where $P,Q \in \mathcal{P}_n$ are $n$-qubit Pauli operators. Thus, $E_{P,Q}(t)$ are the Pauli transfer matrix (PTM) elements~\cite{hashim_practical_2025} of $\widetilde{\mathbbm{1}}(t)$ and we will use this terminology moving forward. 
The time-traces in Eq.~\eqref{eq:PTM_SLT} are the same as those needed for ELT except for the fact that the density matrix $\rho$ has been replaced by a Pauli string $Q$.
These expectation values suffice to reconstruct the Lindblad~\cite{Franca2024Efficient}. 

As the input observable in this case is not a physical state as was the case in ELT, the procedure for estimating the appropriate expectation value $E_i(t)$ cannot involve preparing the state $Q$. Instead, we estimate the expectation values from a classical shadow procedure. 

Experimentally, the protocol for SLT proceeds as follows. 
For each idling time $t$, and for each of the $N_{\text{SLT}}$ independent random instances of the protocol, we do the following procedure, see Fig.~\ref{fig: Figure1}d and Fig.~\ref{fig: Figure2}(b,c):
\begin{enumerate}
    \item \textbf{Precompute model and draw random configurations:} 
    Precompute $\textsf{M}_{(P,Q),j} = \Tr(P\mathcal{L}_j(Q))$ for the Lindblad model of interest $\mathcal{L} = \sum_{j}p_j\mathcal{L}_j$ and find any $\textsf{N}$ for which $\textsf{N}\textsf{M} = \mathbbm{1}$.
    For a number of randomizations, $N_\text{SLT}$, and $N_t$ time steps, $t\in \{t\}$, pick an initialization setting by drawing a Pauli operator and eigenvalue sign uniformly from $\{\pm X,\pm Y,\pm Z\}$ independently for each qubit. Record the eigenvalue signs $\{s_k\}$ and initial states $\{\rho_k\}$ for each qubit, $k$. Select a random Pauli measurement basis by drawing single-qubit measurement bases $B_k$ from $\{X,Y,Z\}$ independently for each qubit.
    \item \textbf{Acquire single-shot measurements:} For each randomization, initialize qubit $k$ in state $\rho_k$ by implementing an initialization pulse $U_k^\text{init}$ and evolve for time $t$ under the idling  dynamics. Measure each qubit in basis $B_k$ by applying a finalization gate $U_k^{\text{meas}}$ and record the outcome signs $\{m_k\}$.
    \item \textbf{Transform and estimate rates by low-polynomial fits:} Estimate $E_{P,Q}(t)$ from the measurement data using classical shadows as described by Eqs.~\eqref{eq:SLT_estimator} and \eqref{eq:SLT-estimator-avg} below for all Pauli operators $P,Q$ necessary to compute the product of Eq.~\eqref{eq:Transformed_PTM_signal}. Transform $\vec{E}$ using Eq.~\eqref{eq:Transformed_PTM_signal} and perform low-polynomial fits over the time axis to extract Lindblad parameters $\Vec{p}$ from the first derivative at $t=0$. In our experiments, we are in the linear-response regime and thus solely utilize linear fits. 
\end{enumerate}
\indent Unlike ELT, which uses a fixed tomographically complete set of probes, SLT reuses the same randomized data to estimate many PTM elements via post-processing. 
In the following section we describe how to estimate $E_{Q,P}(t)$ from the randomized data. 

The goal of SLT is to construct an unbiased estimator for PTM elements in equation~\eqref{eq:PTM_SLT} for any choice of Pauli input string $Q$ and output string $P$.
To do so, we first define a notion of ``agreeing" that helps us to understand the estimator of the shadow protocol. 
Let $P = \bigotimes_{k=1}^n P_k, Q = \bigotimes_{k=1}^n Q_k$ be the tensor-factorization of each Pauli across $n$ qubits, and let $S_1$ and $S_2$ be the set of qubit indices $\{k\}$ for which $P_k\neq I$ and $Q_k\neq I$ respectively. 
Thus, $|S_1|$ and $|S_2|$ are the Pauli weights of $P$ and $Q$ respectively. 
For a given experiment with initial state $\rho = \bigotimes_{k=1}^n \rho_k$ and Pauli measurement operators $B = \otimes_{k=1}^n B_k $ with $B_k\in \{X,Y,Z\}$, we say the shot \emph{agrees} with $P$ if $\mathrm{Tr}(P_k \rho_k) \neq 0$ for all $k \in S_1$, and similarly if $\Tr(Q_kB_k) \neq 0$ for all $k\in S_2$. 
The single-shot estimator of $E_{P,Q}(t)$ for experiment $x$ is then:
\begin{subequations}
\begin{align}
&\text{if $P$ and $Q$ agrees with $x$:}\nonumber\\ 
&\widehat{E}_{P,Q}^x(t) = \Bigg(\prod_{i \in S_1} s_i^x\Bigg)\Bigg(\prod_{j \in S_2} m_j^x\Bigg)\, 3^{|S_1|+|S_2|},\\
&\text{otherwise:}\nonumber\\
&\widehat{E}_{P,Q}^x(t) = 0.
\end{align} \label{eq:SLT_estimator}
\end{subequations}
The intuition is that an experiment will only contribute to the relevant PTM element if it agrees with both its input basis and measurement basis and should be put to $0$ otherwise. 
The product of the state initialization signs, $s_i$, and the measurement signs, $m_i$, represent whether the experiment gave rise to a total $+1$ or $-1$ contribution of the expectation value, and the constant factor represents a normalization with how rarely an experiment will agree with the Pauli strings $P$ and $Q$.

After performing $N_\text{SLT}$ experiments, the resulting estimate of $E_{P,Q}(t)$ is then obtained from the average:
\begin{equation}\label{eq:SLT-estimator-avg}
    \widehat{E}_{P,Q}(t) = \frac{1}{N_{\text{SLT}}}\sum_{x=1}^{N_{\text{SLT}}}\widehat{E}_{P,Q}^{x}(t)
\end{equation}

In shadow tomography, it is necessary to show that the estimator is unbiased and has bounded variance.
It has been shown that the estimator in Eq.~\eqref{eq:SLT_estimator} is unbiased, with variance bounded by~\cite{Franca2024Efficient}
\begin{align}
    \mathbbm{E}\left(\widehat{E}_{P,Q}(t)\right) &= \frac{1}{2^n}\Tr(P\widetilde{\mathbbm{1}}(t)(Q)),\\
    \text{Var}\left(\widehat{E}_{P,Q}(t)\right) &\leq 9^{|S_1| + |S_2|}. \label{eq:SLT_Franca_var}
\end{align}
In addition, it is shown that for a $k$-local Lindblad the weight of Pauli strings required for reconstruction does not scale with system size. In this case, we should therefore think of $|S_1| + |S_2|$ as small which yields a variance of constant size.  

Typically in shadow tomography, the median-of-means statistical approach is used to limit the effects of outliers \cite{Huang_shadows_2020,Franca2024Efficient, ShadowQST, chen_robust_2021}.
This approach ensures that the estimator across many experiments will get close to the value of interest with high probability, despite the potential for heavy-tailed outcome distributions \cite{Huang_shadows_2020}. 
However, for the bounded-$k$ regime, we now show that a traditional Gaussian statistics approach is sufficient for estimating PTM elements with high probability.

The estimator of Eq.~\eqref{eq:SLT_estimator} is bounded by $|\widehat{E}_{P,Q}| \leq 3^{|S_1|+|S_2|}$ for every experiment.
Given this property, we construct the empirical mean-estimate of Eq. \eqref{eq:SLT-estimator-avg}.
Application of Hoeffding's inequality results in strong concentration around the parameter of interest.
In particular, we find
\begin{align}
    \textrm{Prob}(|E_{P,Q}-\widehat{E}_{P,Q}|\geq  \varepsilon) \leq 2\exp(-\frac{\varepsilon^2N_{\text{SLT}}}{2\cdot  9^{|S_1| + |S_2|}}),
\end{align} 
which thus provides exponentially good confidence for empirical means for our case of low-weight Pauli operators $|S_1| + |S_2| = O(1)$.

Concretely, this means we can estimate exponentially many low-weight observables, without resorting to median-of-means estimators.
This property is crucial; it makes SLT compatible with standard error propagation and $\chi^2$-regression, unlike many shadow protocols where heavy-tailed estimators complicate uncertainty quantification.
In the rest of this work, we use this attribute to estimate errors from low-polynomial $\chi^2$-fits and propagate them through the Rice distribution to get uncertainties on the magnitude of extracted parameters, analogous to ELT. 

Repeating the randomized acquisition of $\vec{E}(t)$ for the set of short evolution times $\{t\}$ yields time traces analogous to those in ELT, see Fig.~\ref{fig: Figure2}(b). 
Analogously to ELT, these traces are then transformed into $P(t)$ according to Eq.~\eqref{eq:Transformed_PTM_signal}, and the derivative at $t=0$ is extracted via low-order polynomial fits to recover the Lindblad parameters $\vec{p}$, see Fig.~\ref{fig: Figure2}(c).

Similar to ELT, assuming at most on-site dissipation and $k$-local coherent dynamics results in each parameter only depending on a few low-weight PTM elements. 
As the shadow estimation procedure provides exponentially good confidence for low-weight PTM elements, this implies that the necessary number of randomizations, $N_\text{SLT}$, scale logarithmically with system size, see Fig.~\ref{fig: Figure1}(b). 
As such, SLT is preferable to ELT in the physically motivated regime of systems where interactions are at most $k$-body and $k$ does not scale with system size. 
When assuming nothing about the locality structure of the underlying LME, SLT could in general perform worse, as the variance of each expectation value might be exponentially larger in the worst case. 
For any physical system, this motivates the question whether it is in the physically motivated regime in which SLT outperforms ELT with the same number of shots or in the poorly performing regime in which SLT is unable to estimate any parameter with good confidence. 
In the following sections we will investigate this question on our 5-qubit processor by benchmarking the parameters found in SLT against ELT as a function of the number of randomizations.

\section{Single qubit verification of extensible and shadow protocols}\label{sec:1qubit}

\begin{table}[t]
\centering
\label{tab:exp_configs}
\renewcommand{\arraystretch}{1.2}
\begin{tabular}{c|p{3cm}|p{3.4cm}}
\hline
Qubits & \multicolumn{1}{c|}{ELT} & \multicolumn{1}{c}{SLT} \\
\hline
$n=1$ & $N_\text{conf}=18$ \newline $N_\text{shot}=500$ \newline $N_{\text{ELT}} = 9000$ \newline $t_{\text{max}} = 30 \mu s$ & $N_{\text{SLT}}\in \{10,30,100,\newline300,1000, 3000,9000\}$ \newline $t_{\text{max}} = 30 \mu s$ \\
\hline
$n=3$ & $N_\text{conf}=5832$ \newline $N_\text{shot}=2000$ \newline $N_{\text{ELT}} = \num{11664000}$ \newline $t_{\text{max}} = 1 \mu s$& $N_{\text{SLT}}\in\{10,10^2,10^3,\newline 10^4, 10^5, 10^6,10^7\}$ \newline $t_{\text{max}} = 1 \mu s$  \\
\hline
$n=5$ & n/a & $N_{\text{SLT}} = *\num{25000000}*$ \newline $t_{\text{max}} = 3 \mu s$ \\
\hline
\end{tabular}
\caption{Experimental configurations used for ELT and SLT. ELT uses fixed configuration sets of size $N_\text{conf}$ with $N_\text{shots}$ repeated shots for a total number of experimental shots $N_\text{ELT}=N_\text{conf}N_\text{shot}$, where SLT subsamples $N_{\text{SLT}}$ of these configurations for each time step. Additionally, both procedures are implemented with $N_t=20$ time steps uniformly spaced in the range $[0 \mu s,t_{\text{max}}]$ with varying $t_\text{max}$ for the different experiments. The stars in the table is to make aware that for the five qubit experiment, Sec.\ \ref{sec:5qubits}, we attempt to learn the parameters of a 2-local model. In the single and three qubit experiments, Secs.\ \ref{sec:1qubit} and \ref{sec:3qubits}, we set out to learn all Lindblad parameters.}
\end{table}

\begin{figure}[t!]
    \centering
    \includegraphics{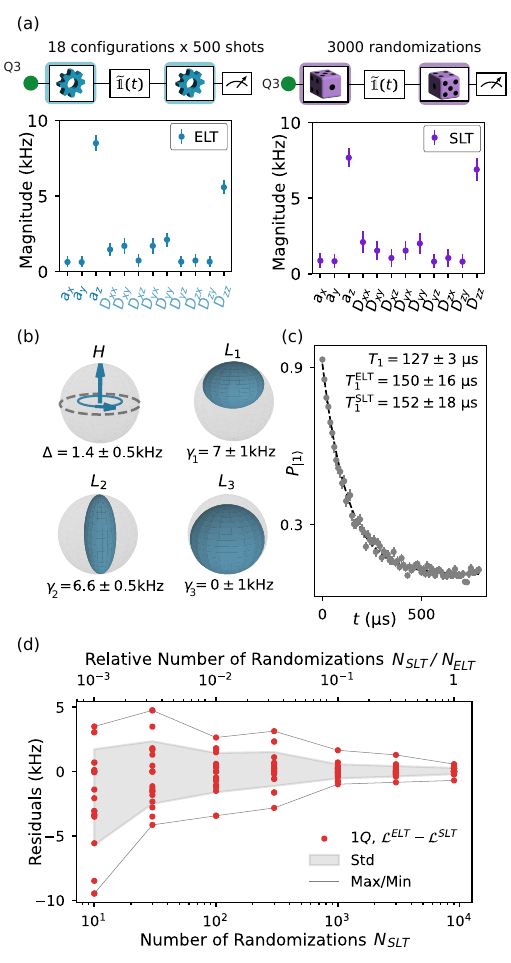}
    \caption{ELT and SLT on a single qubit. 
    \textbf{(a)} Magnitude of the estimated parameters from both the ELT and SLT procedure.
    \textbf{(b)} A visual representation of the fitted Lindblad parameters from ELT. The ground state of the Hamiltonian is shown on the Bloch sphere and the three extracted Lindblad jumping operators are represented by how they deform the Bloch sphere, along with their respective rates. 
    \textbf{(c)} A subset of the data used for ELT/SLT can be interpreted as a $T_1$-experiment. 
    The data and a traditional $T_1$ experiment is shown together with the extracted $T_1$ from ELT and SLT. 
    \textbf{(d)} Difference in all parameters extracted from the two methods (in red points) plotted against the number of random instances used to do SLT.}
    \label{fig: Figure3}
\end{figure}

We first benchmark ELT and SLT on a single qubit to validate the protocol. We do so by comparing ELT to independently calibrated device parameters and by comparing the extracted of SLT to the ones extracted from ELT.
For both procedures, we use $N_t=20$ time steps uniformly spaced between 0 and $30\mu$s. For ELT, we perform $N_{\text{shot}}=500$ shots for each of the six input Pauli states and three Pauli measurement bases, for a total of $N_\text{conf}=18$ input--output configurations.
For SLT, we use randomization sizes in the set $N_{\text{SLT}}\in \{10,30,100,300,1000,3000,9000\}$ (see Table~\ref{tab:exp_configs}), and unless otherwise stated we highlight results for $N_\text{SLT}=3000$ randomizations.
To ensure a fair comparison and avoid effects due to parameter drift of the device, SLT randomizations are obtained by subsampling the ELT data set with no repetitions.
 
In Fig.~\ref{fig: Figure3}(a), we show the magnitudes of all 12 on-site parameters extracted from ELT and from SLT.
The dominant extracted parameters are $a_z$ and $D_{zz}$, corresponding to detuning and dephasing of the qubit, consistent with known noise sources in superconducting qubits~\cite{engineers_guide_morten_19}.
All parameters extracted from the two methods agree within their uncertainties.
 
To verify the physical plausibility of the reconstructed Lindblad we apply two independent checks, shown in Fig.~\ref{fig: Figure3}(b,c).
In Fig.~\ref{fig: Figure3}(b) we visualize the 12 parameters extracted from ELT on the Bloch sphere.
The first sphere shows the ground state of the extracted Hamiltonian together with its energy splitting $\Delta$ in the rotating frame of the experimental drive qubit frequency.
We then diagonalize the dissipation matrix $D$ to obtain three Lindblad jump operators with associated rates, and show in simulation how each of them deforms an initial Bloch sphere after a fixed evolution time.
The Hamiltonian closely matches that of a slightly detuned drive frequency compared to the qubit transition frequency, and the three extracted jump operators closely resemble textbook relaxation, dephasing and temperature-induced excitation channels \cite{nielsen_and_chuang}.
 
Second, we compare ELT and SLT against a standard $T_1$ measurement.
In Fig.~\ref{fig: Figure3}(c) we extend the ELT idling time to $800\,\mu\mathrm{s}$.
In this limit, the ELT setting that prepares $\ket{1}$ and measures in the $Z$ basis is exactly a conventional $T_1$ decay experiment.
We therefore fit $T_1$ directly from this trace and compare it to the $T_1$ inferred from the jump operators reconstructed by ELT and SLT.
The resulting values agree within measurement uncertainty (1.5 standard deviations), supporting the use of ELT as our reference baseline in what follows.
As expected, the dedicated $T_1$ fit achieves substantially smaller uncertainty.
It is optimized for estimating a single parameter, whereas ELT and SLT are designed for broadly model-agnostic reconstruction rather than high-precision extraction of any one decay channel.

Finally, we benchmark SLT against ELT and study how the SLT reconstruction converges with the number of randomizations.
Figure~\ref{fig: Figure3}(d) shows the parameter-wise residuals (subtracting SLT from ELT parameters).
For each subsampling level, we plot the individual residuals together with their standard deviation and overall min/max range.
As the number of randomizations increases, the two protocols become increasingly consistent.
At $N_\text{SLT} \sim 10^3$ randomizations, the typical residual reaches the $\sim\mathrm{kHz}$ level, after which the residual spread saturates.
This saturation is consistent with the average ELT parameter of $\sigma_{\text{ELT}} = 0.5$ kHz uncertainty setting the floor; if SLT shot noise were the dominant contribution, we would instead expect central limit-like scaling $\sigma \propto 1/\sqrt{N_\text{SLT}}$.
Taken together, this indicates that SLT might reduce experimental cost without sacrificing accuracy. In the next section, we investigate how the experimental cost of ELT and SLT scales as we increase the system size to 3 qubits.

\section{Full reconstruction of three qubit dynamics and benchmark of shadows}\label{sec:3qubits}
\begin{figure}[t!]
    \centering
    \includegraphics{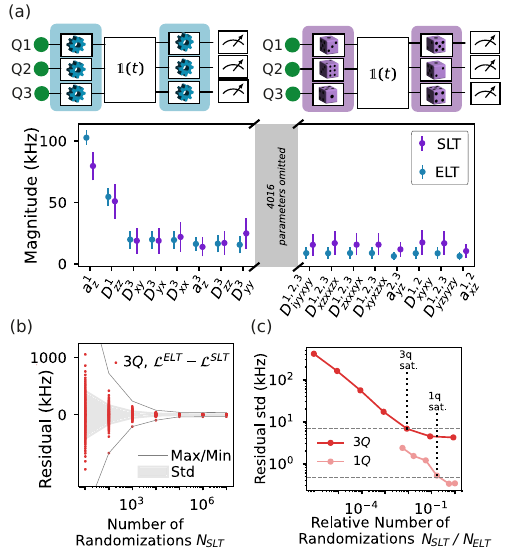}
    \caption{Extensible Lindblad tomography and shadow Lindblad tomography for three qubits. 
    \textbf{(a)} The magnitude of the fitted Lindblad parameters both for ELT and SLT ordered by their size estimated in ELT. 
    We have here omitted 4016 parameters for clarity. 
    Error bars are 16 and 84 percent quantiles of the magnitude. 
    \textbf{(b)} Difference in extracted parameters of the two methods (in red points) plotted against the number of shadows used to do SLT. 
    The plot is capped to omit the largest deviating residuals for clarity. \textbf{(c)} The standard deviation of the residual plotted against the relative size of the number of shadows used for SLT compared to the full number of shots done in ELT for both a single qubit and 3 qubits. The dashed horizontal lines show the average uncertainty from ELT on three qubits and a single qubit respectively. Both residual standard deviations go down with the number of randomizations before saturating around the horizontal lines in the plot. We see that the saturation point is relatively lower for 3 qubits than for a single qubit, a property which is consistent with being in the $k$-local regime.}
    \label{fig: Figure4}
\end{figure}

Building on the single-qubit validation, we now benchmark SLT against ELT on three qubits simultaneously on our 5-qubit processor.
For ELT we implement all $6^3$ input Pauli states and all $3^3$ Pauli measurement bases ($N_\text{conf}=5832$), let the system evolve for up to $t_\text{max}=1\mu\mathrm{s}$ and acquire $N_{\text{shot}}=2000$ shots per setting.
SLT reuses the ELT data via subsampling, with the number of randomizations $N_{\text{SLT}}$ varying from $10$ to $10^7$ in logarithmically spaced steps (see Table~\ref{tab:exp_configs}).
At this point we make no structural assumptions about the Lindblad generator and estimate all $4^3(4^3-1)=4032$ 3-qubit parameters of the LME in Pauli form.

Figure~\ref{fig: Figure4}(a) shows the magnitude of extracted parameters ordered according to the magnitude found in ELT.
The largest extracted parameters are all single-qubit terms, consistent with standard superconducting qubit properties~\cite{engineers_guide_morten_19}.
In particular, the dominant contributions correspond to detuning ($a_z$), dephasing ($D_{zz}$), and relaxation channels ($D_{xx}, D_{xy}, D_{yx}, D_{yy}$), in line with other reports of Lindblad tomography on superconducting processors~\cite{samach2022lindblad}.

To assess SLT performance, Fig.~\ref{fig: Figure4}(b) shows the residuals estimates (subtracting SLT from ELT parameters) as a function of the SLT randomization count, similar to Figure~\ref{fig: Figure3}(d).
The standard deviation of the residual decreases with the number of randomizations, as expected for statistical convergence.
A subset of around $10^5$ randomizations suffices for ELT and SLT to agree at the level of a few kilohertz. 

Finally, Fig.~\ref{fig: Figure4}(c) compares the standard deviation of residuals across one-, and 3-qubit experiments as a function of the fraction of SLT samples relative to the total number of ELT shots.
Over a wide range of sample sizes, the standard deviation follows an expected $\sigma \propto 1/\sqrt{N}$ trend, but both curves saturate to around the average uncertainty of the ELT parameters at a fraction of the full ELT dataset.
This saturation indicates that, beyond this point, the uncertainty of the residuals is dominated by the ELT uncertainties, so increasing the SLT sample size no longer improves agreement.
In this regime SLT is therefore both more shot-efficient and at least as accurate as ELT. 

We find that  every 3-local coherent and incoherent parameter found is less than one standard deviation away from 0. This suggests that, within our experimental uncertainty, full agnostic Lindblad characterization is overkill, and restricting a model to 2-local dynamics would be sufficient for finding all non-zero parameters.

Another signature verifying this conclusion is the observation that saturation occurs at a smaller SLT sample fraction for three qubits than for a single qubit.
Should the dynamics be well-described by 1- or 2-local interactions, only PTM elements of the idling channel with lower weight than three should be non-zero for short enough times. With SLT, we can estimate these non-zero expectation values with fewer shots than would be necessary in ELT for a given target precision. Thus, it would then be expected that the 3-qubit SLT procedure saturates at a lower relative shot count than single qubit SLT, as there is no sampling-theoretical benefit on the single qubit level.

The earlier saturation of the residual standard deviation is therefore consistent with a $k$-local interaction assumption. As this implies $\log(n)$ scaling of the protocol to higher system sizes, this reinforces that shadow tomography can enable parameter learning for large-scale systems under realistic noise.

\section{Five qubit Shadow Lindblad tomography under local interactions}\label{sec:5qubits}
Fully unrestricted Lindblad reconstruction in the Pauli basis quickly becomes impractical beyond three qubits; estimating all $4^5(4^5-1)=\num{1047552}$ parameters for five qubits via ELT would require roughly 162 days of data acquisition under our experimental conditions. 
Having validated SLT against ELT for single- and 3-qubit systems, and found 3-local dynamics to be negligible, we now adopt a physically motivated restriction on the Lindblad model to achieve scalability.
Specifically, in what follows we assume that the dynamics are described by on-site dissipation and pairwise coherent interactions only.
This corresponds to the standard modeling of superconducting-qubit processors, where local Lindblad terms set coherence times, on-site Hamiltonian terms generate single-qubit dynamics, and two-body interactions capture residual entangling evolution.

\begin{figure}[t!]
    \centering
    \includegraphics[width=\linewidth]{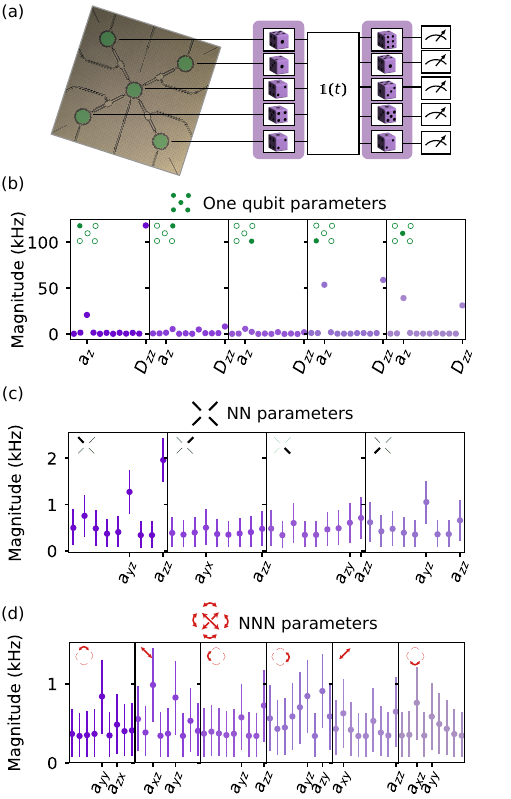}
    \caption{SLT on a 5-qubit processor. \textbf{(a)} Characterizing the Lindblad dynamics for the entire five qubit device assuming only single qubit loss and two-qubit interactions.
    \textbf{(b)} The magnitude of extracted on-site Lindblad parameters. The x-axis is the same as in Fig.~\ref{fig: Figure3}(a). \textbf{(c), (d)} The magnitude of all extracted nearest neighbor and next-nearest neighbor interactions. The x-ticks are $a_{ij}$ with $i,j \in \{x,y,z\}$ in lexicographic order.}
    \label{fig5:Figure5}
\end{figure}

Within this restricted model, we perform SLT on the full 5-qubit processor.
We implement $N_{\text{SLT}} = \num{25000000}$ randomizations for $N_t=20$ evolution times up to $t=3\mu\mathrm{s}$ for a total acquisition time of around 9 hours. 
Figures~\ref{fig5:Figure5}(b,c,d) report the magnitude of the extracted on-site qubit parameters, nearest-neighbour interactions and next-nearest-neighbour interactions, respectively.
Comparing Fig.~\ref{fig5:Figure5}(b) to Figs.~\ref{fig5:Figure5}(c,d), we see that the single-qubit parameters are typically two orders of magnitude larger than the two-qubit parameters, consistent with expectations for weakly coupled transmon arrays.
Moreover, with the tunable couplers idled during acquisition, the extracted interaction strengths remain bounded to at most a few kilohertz, confirming that residual static coupling is minimal across the 5-qubit processor.

We estimate, that doing the same experiment for ELT in a restricted model setting would take $N_{\text{ELT}} = \num{160000000}$ experiments or take 57 hours of data acquisition. Accordingly, our 5‑qubit SLT experiment reaches a parameter regime that is prohibitive for unrestricted ELT and remains resource‑intensive even under a 2‑local ELT assumption. These results position SLT as a scalable primitive for Lindblad tomography on larger processors. 

\section{Discussion and Outlook}
Here we summarize the main takeaways and outline near-term directions.

A strength of our shadow procedure is that, in contrast to other existing shadow tomography schemes, the constructed estimator not only has bounded variance, but is bounded for any single-shot experiment. 
As a consequence, the estimator is fully compatible with ordinary error propagation, enabling transparent uncertainty quantification on extracted parameters. This capability makes SLT particularly valuable as a benchmark for analog quantum computation, allowing expectation values from analog quantum simulations to be reported with statistically rigorous confidence intervals, e.g., as in Ref.~\cite{zoller_bounded-error_2025}.

SLT complements GST \cite{nielsen_gate_2021} and randomized benchmarking (RB)~\cite{hashim_practical_2025}.
RB reports coarse-grained error rates and GST targets gate-level noise models, whereas SLT targets idling dynamics and yields physically interpretable Lindblad coefficients with built-in error bars and efficient reuse of randomized data.
In analog quantum computing regimes, SLT therefore finds particular application due to its ability to learn Lindblad parameters in an efficient manner.

Both ELT and SLT as implemented here assume time-independent, Markovian dynamics described by a Lindblad generator.
Recently, this scheme has been extended to a time-dependent setting~\cite{Franca2025_time_dependent}, which would enable direct characterization of explicitly time-dependent processes such as driven gates within the same SLT workflow.

A shared limitation of ELT and SLT is sensitivity to state-preparation and measurement (SPAM) errors, since the prepared inputs and measurement bases serve as probes of the dynamics.
Following Ref.~\cite{samach2022lindblad}, one can calibrate the effective states and POVMs from the $t=0$ setting and use this to map measured expectation values to corrected ones.
However, such a characterization is only possible up to a choice of gauge \cite{nielsen2021gate}, and so further physically motivated assumptions would be necessary for full SPAM-mitigation. We outline a concrete mitigation scheme in Appendix~\ref{app:spam}, but here we instead prioritized matching ELT/SLT comparisons under shared experimental conditions and did not implement full SPAM mitigation.

To quantify the SPAM-error and to verify that probe preparations are sufficiently accurate for ELT and SLT, we characterize all five qubits (Appendix~\ref{app:singlequbitproperties}). Using interleaved RB \cite{hashim_practical_2025} (both sequential and parallel) and state-discrimination, we observe gate fidelities $\geq 99.8\%$ and assignment fidelities $\gtrsim 90\%$. We therefore conclude that gate errors are negligible in our setting, while SPAM errors constitute a dominant bottleneck for precise estimation of Lindblad parameters. While this constrains absolute accuracy, it does not affect our comparative conclusions on SLT and ELT.

Another relevant error source is leakage out of the computational subspace.
With two-level discrimination, leakage population is misassigned as $\ket{0}$ or $\ket{1}$ and cannot be corrected without additional characterization.
However, we note that since ELT and SLT inversion is linear in the estimated expectation values, a readout fidelity $\gtrsim 90\%$ implies that leakage-induced misassignment can be expected to result in up to 10\% biases in inferred Lindblad parameters. 
Future experiments should incorporate explicit leakage diagnostics (e.g., leakage RB \cite{wallman_robust_2016}) to quantify and correct these effects.

\section{Conclusion}
This work tests the central premise that shadow Lindblad tomography (SLT) can outperform extensible Lindblad tomography (ELT) in experimentally relevant settings. Across one‑ and three‑qubit experiments, SLT agrees with ELT within uncertainties while using substantially fewer measurements, and the relative data fraction required by SLT decreases with system size, consistent with behavior of $k$-local models, indicating a possible $\log(n)$ resource scaling. In a five‑qubit demonstration under a restricted Lindblad model, we fully reconstruct on‑site dissipative and 2‑local coherent parameters, finding single‑qubit terms approximately two orders of magnitude larger than two‑qubit terms—consistent with expectations for our superconducting hardware and in phenomenological agreement with parameters extracted from single‑qubit ELT.
Finally, we establish compatibility with standard $\chi^2$-fitting to emphasize transparency and reproducibility. Using Hoeffding-type argument, we show that SLT estimators have Gaussian uncertainties and by direct experimental test, we verify that $\chi^2$-procedures yield statistically consistent parameter estimates with previous robust fitting procedures~\cite{Franca2024Efficient}.
This simplification avoids median-of-means to propagate errors and eases practical deployment of SLT.

These results motivate scaling SLT to larger systems. E.g., for 50 qubits, under a $2$‑local model, learning all couplings and on‑site coherent and incoherent parameters to kHz precision is estimated to require $\sim$5 years of data acquisition with ELT versus $\sim$22 hours with SLT, positioning SLT as an efficient subroutine for characterizing idling noise—even in full‑scale superconducting processors.

\begin{acknowledgments}
This research was supported by the Novo Nordisk Foundation (grant no. NNF22SA0081175), the NNF Quantum Computing Programme (NQCP), Villum Foundation through a Villum Young Investigator grant (grant no. 37467, grant No. 25452 and grant No. 60842), the QMATH Centre of Excellence (grant No. 10059), the Innovation Fund Denmark (grant no. 2081-00013B, DanQ), the U.S. Army Research Office (grant no. W911NF-22-1-0042, NHyDTech), by the European Union through an ERC Starting Grant, (grant no. 101077479, NovADePro) and by the Carlsberg Foundation (grant no. CF21-0343). Any opinions, findings, conclusions or recommendations expressed in this material are those of the author(s) and do not necessarily reflect the views of Army Research Office or the US Government. We acknowledge financial support from the Knut and Alice Wallenberg Foundation (KAW) through the Wallenberg Center for Quantum Technology (WACQT).  The device was fabricated at the Myfab Chalmers Nanofabrication Laboratory.

Views and opinions expressed are those of the author(s) only and do not necessarily reflect those of the European Union or the European Research Council. Neither the European Union nor the granting authority can be held responsible for them. 
Finally, we gratefully acknowledge Lena Jacobsen for program management support.
\end{acknowledgments}

\newpage
\appendix
\onecolumngrid

\setcounter{figure}{0}
\renewcommand{\thefigure}{S\arabic{figure}} 
\setcounter{table}{0}
\renewcommand{\thetable}{S\arabic{table}}
\setcounter{equation}{0}
\renewcommand{\theequation}{S\arabic{equation}}
\renewcommand{\theHfigure}{S\arabic{figure}}
\renewcommand{\theHtable}{S\arabic{table}}
\renewcommand{\theHequation}{S\arabic{equation}}

\section{Hardware setup and device architecture}\label{app:setup}
\begin{figure}[h!]
    \centering
    \includegraphics[width = 0.8\textheight]{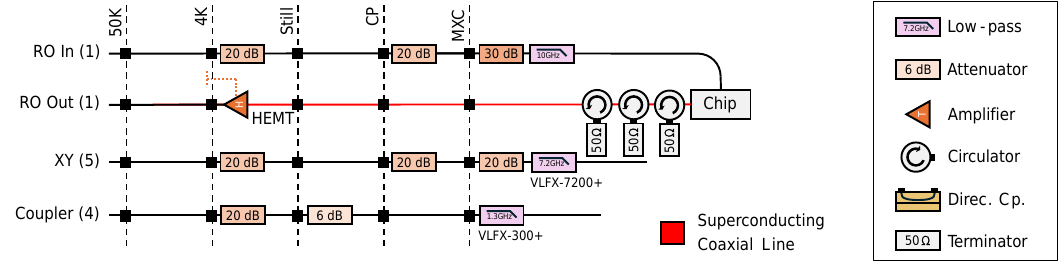}
    \caption{The cryogenic setup used for the experiments.}
    \label{fig:setup}
\end{figure}

In this section we provide an overview of the experimental setup. 
The device used in this work was designed and fabricated at Chalmers University of Technology and is a five qubit processor with fixed frequency transmons and tunable frequency couplers.
A lithographically identical device to the one shown in Figure~\ref{fig: Figure1}(a) is mounted in a QCage.24 and loaded in a BlueFors XLD1000.
For pulse control and qubit readout, we use high frequency control units in Quantum Machines OPX1000.
All experiments are orchestrated with our in-house python data acquisition platform called \textit{Pelagic}.
The cryogenic setup is summarized in Fig.~\ref{fig:setup}.

Additional details of device fabrication may be found in Ref. \cite{biznarova_mitigation_2024}.

\section{SPAM considerations and mitigation}\label{app:spam}

In this work, we do not apply explicit State Preparation and Measurement (SPAM) mitigation during analysis. Because our inputs are Pauli expectation values that are subsequently transformed into transformed PTMs via Eq.~\eqref{eq:Transformed_PTM_signal}, any bias in the measured expectations propagates linearly into the recovered Lindblad parameters. This sensitivity is inherent: preparation and measurement act as dynamical probes, so imperfections in either channel directly affect slopes at the intercept.

The snapshot at $t=0$ provides partial information about SPAM. For each probe $(\rho_i, O_i)$, the $t=0$ data point amounts to a (basis-restricted) calibration of the prepared state and measured POVM along $O_i$, i.e., a partial tomography in the chosen measurement basis. In principle, these data can be used to infer corrections to the expectation values prior to transformation of Eq.~\eqref{eq:Transformed_PTM_t=0}. However, this reconstruction is only defined up to a gauge: without additional constraints, simultaneous identification of both the faulty preparation and faulty POVM is underdetermined \cite{nielsen2021gate}. As a result, full SPAM mitigation requires physically motivated assumptions to fix the gauge.

A practical path is to adopt a physics-informed SPAM model. For superconducting transmons, preparation errors can be approximated by residual thermal population in the initialized state, while measurement errors are well described by per-qubit confusion matrices (classical mixtures of correct classification and false positives/negatives). When these confusion matrices are well-conditioned, one can invert them to deconvolve measurement errors and obtain corrected expectation values prior to forming transformed PTMs. Noise mitigating faulty measurement in this sense is experiment-specific and can be driven by $t=0$ calibration records in the same measurement bases. Correcting state-preparation errors is harder: Once the system evolves under $\widetilde{mathbbm{1}}(t)$, the initial-state bias is irretrievably mixed into the dynamics and cannot be uniquely disentangled without an explicit Lindblad model for $\rho_i$ (e.g., a thermal-population parameter per qubit) that is jointly fit with the Lindblad parameters. Absent such a model, preparation biases should be treated as nuisance parameters, and their impact on the slopes should be bounded. 

In practice, we prioritize matched comparisons between ELT and SLT under identical measurement chains and use readout-fidelity bounds to contextualize accuracy. With two-outcome discrimination fidelities $\gtrsim 90\%$, the resulting expectation-value biases are at the $\sim 10\%$ level, which informs absolute accuracy but does not compromise the comparative conclusion that ELT and SLT agree under shared SPAM channels. Future deployments can incorporate (i) routine $t=0$ POVM calibration and inversion in the experiment’s native bases and (ii) simple thermal-preparation models with joint fitting or priors. These steps provide a practical mitigation path and maintain transparency in uncertainty quantification.

\section{Single qubit characterization experiments}\label{app:singlequbitproperties}
\begin{figure}[t!]
    \centering
    \includegraphics[width=\linewidth]{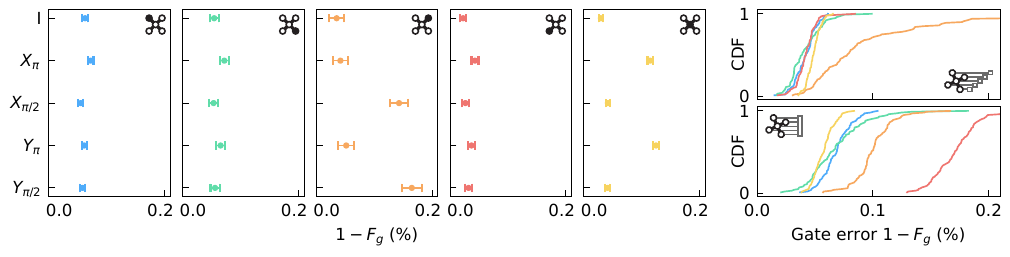}
    \caption{Single qubit interleaved randomized Clifford benchmarking error rates for the gate set $\{I, X_\pi, X_{\pi/2}, Y_\pi, Y_{\pi/2}\}$ on all five qubits. 
    We find IRB error rates consistently below $0.2\%$. 
    In the right panel we show the CDF's of CRB-inferred error rates of 15 minutes of repeated benchmarking in series (top) and simultaneous (bottom).}
    \label{fig: IRB}
\end{figure}

\begin{figure}[h!]
    \centering
    \includegraphics[width=\linewidth]{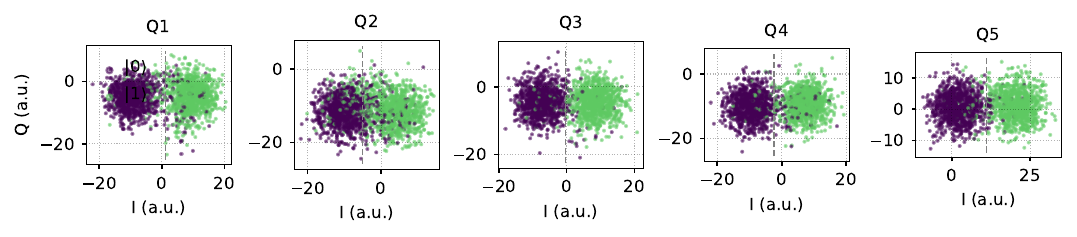}
    \caption{IQ measurements for all five qubits initialized in the $\ket{0}$ and $\ket{1}$, respectively. We consistently get readout fidelities of $\gtrsim 90\%$, which is also the case for all qubits in this figure, except for qubit 2.}
    \label{fig: IQ}
\end{figure}

In order to characterize the single qubit performance, we perform two experiments: Interleaved Randomized Benchmarking to get a feeling of the error rates of all gates and IQ-thresholding to estimate the false positive and false negative error rate of the qubits. 
We do so, as ELT and SLT needs controllable initial state preparation, corresponding to initial single-qubit pulses and a high-fidelity measurement, corresponding to final single-qubit pulses and a state-discrimination task. 
In Figure~\ref{fig: IRB}, we show interleaved single-qubit randomized benchmarking on each of the 5 qubits sequentially.
The error rates are in general found to be below $0.2\%$, indicative that the single-qubit gates are contributing errors at a level negligible for the main conclusions in this work.
In the right panel of Fig.~\ref{fig: IRB} we compare error rates extracted from sequential (top) and simultaneous (bottom) interleaved single-qubit RB.
The distribution of error rates (plotted as CDF) corresponds to 15 minutes of repeated benchmarking.
In Figure \ref{fig: IQ}, we show a representative measurement of the IQ-blobs for a  measurement in respectively the $\ket{0}$ and the $\ket{1}$-state. Before any experiment, we optimized measurement fidelity, and consistently had measurement fidelity of $\gtrsim 90\%$. 

\section{Fitting Procedure (robust vs.~$\chi^2$)}\label{app:fitting}

França et\,al.~\cite{Franca2024Efficient} propose a robust low-polynomial fitting scheme designed to estimate short-time derivatives with statistical guarantees under heavy-tailed or contaminated measurements. For completeness, we summarize their method below and experimentally compare extracted Lindblad parameters from this fitting method to ordinary $\chi^2$ regression on two-qubit ELT data.

There are three things that should be noted before the polynomial fitting procedure: What the error  and fitting models are, which times we should sample the signal and which loss-functions we fit against. 
\paragraph{Model and contamination assumption.}
Consider a time series $\{(t_i, y_i)\}_{i=1}^T$ of transformed PTMs, where $y_i$ denotes the transformed expectation at time $t_i$. Assume that at least half of the samples satisfy
\[
|y_i - \bar{y}_i| \le \sigma, \qquad \bar{y}_i = \sum_{k=0}^{K} p_k\, t_i^k,
\]
for some (unknown) polynomial coefficients $\{p_k\}$ and contamination level $\sigma$. The goal is to recover the slope at the intercept, i.e., the derivative at $t=0$ of the underlying polynomial.

\paragraph{Time selection and Chebyshev measure.}
Let $t_{\min}, t_{\max}$ denote the time interval used for short-time estimation and define $\bar{t} \equiv (t_{\max}+t_{\min})/2$ and $t_{\Delta} \equiv (t_{\max}-t_{\min})/2$. The robust approach samples times $\{t_i\}_{i=1}^T$ according to the Chebyshev measure mapped to $[t_{\min}, t_{\max}]$. That is, we sample times according to the density
\begin{equation}
    \rho(t) = \frac{t_\Delta}{\sqrt{t_{\Delta}^2 - (t-\bar{t})^2}},
\end{equation}
in order to stabilize polynomial estimation over the interval \cite{Franca2024Efficient}. 
In our experiments, we have instead use uniformly spaced times for feasibility and hardware simplicity.

\paragraph{Chebyshev partitions and losses.}
Partition $[t_{\min}, t_{\max}]$ into $M$ Chebyshev subintervals, i.e. construc the intervals::
\begin{equation}
    I_j = \Big[t_{\Delta}\cos\!\Big(\frac{\pi j}{M}\Big) + \bar{t},\;\; t_{\Delta}\cos\!\Big(\frac{\pi (j-1)}{M}\Big) + \bar{t}\Big], \quad 1 \le j \le M.
\end{equation}
Define two losses for a candidate polynomial $p(\cdot)$:
\begin{align}
    L_1(p) =& \sum_{j=1}^{M} |I_j|\, \mathrm{mean}_{\,t_i \in I_j}\big|y_i - p(t_i)\big|,
\\
L_{\infty}(p) =& \max_{1 \le j \le M} \big| p(\tilde{t}_j) - \tilde{y}_j \big|,
\end{align}
where $\tilde{t}_j \in I_j$ is any representative point and $\tilde{y}_j \equiv \mathrm{median}_{\,t_i \in I_j} y_i$ is the median of the data in the interval.
With this settled, we can introduce the fitting algorithm itself. 

\paragraph{Robust fitting algorithm.}
The robust procedure proceeds in two phases:
\begin{enumerate}
    \item \textbf{Baseline fit ($L_1$):} Find a polynomial $p$ that minimizes $L_1(p)$; form residuals $z_i = y_i - p(t_i)$.
    \item \textbf{Iterative corrections ($L_{\infty}$):} For $k=1,2,\dots$, fit a correction polynomial $q_k$ by minimizing $L_{\infty}(q_k)$ over the residuals, update $z_i \leftarrow z_i - q_k(t_i)$, and stop when $\max_j |q_k(\tilde{t}_j)|$ falls below a predefined threshold.
\end{enumerate}
The final robust estimate is then the polynomial
\begin{equation}
    p_{\mathrm{tot}}(t) \;=\; p(t) + \sum_{k=1}^{k_{\mathrm{end}}} q_k(t).
\end{equation}
This procedure can be efficiently implemented, as both steps can be formulated as a linear program.  

\begin{figure}[t!]
    \centering
    \includegraphics[width=0.7\linewidth]{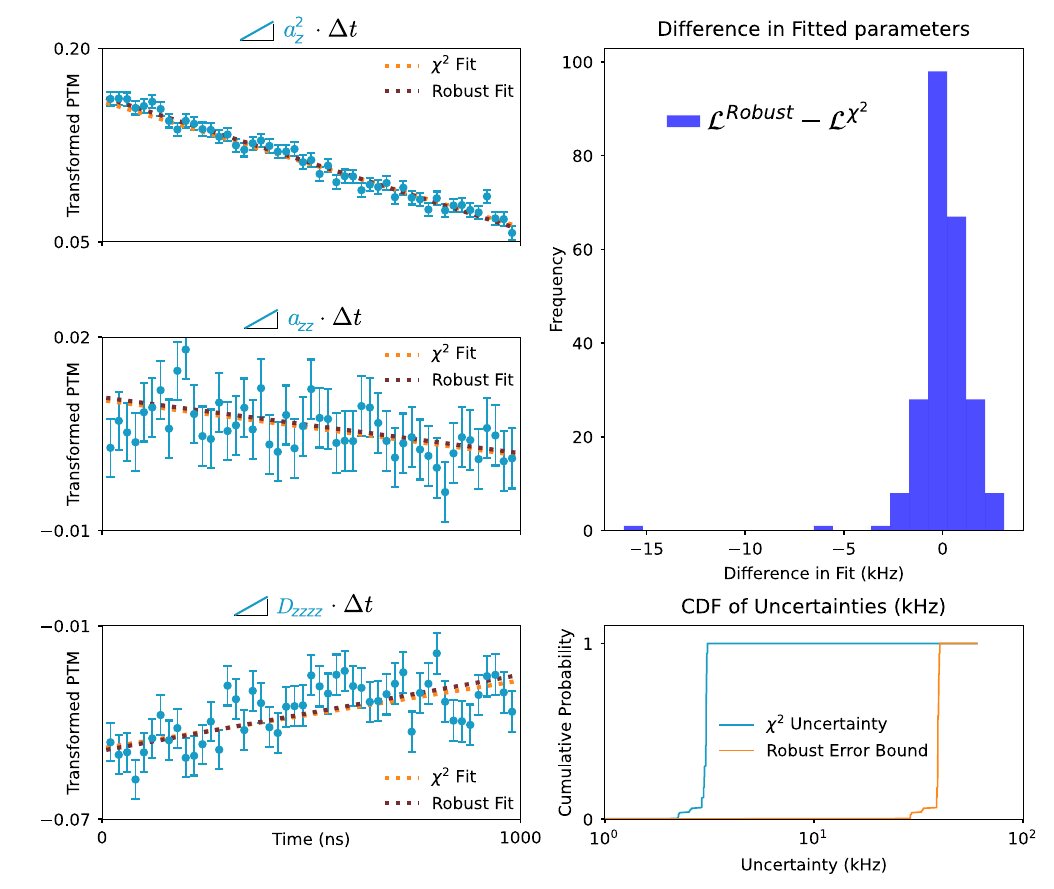}
    \caption{Comparison of $\chi^2$ regression and robust low-polynomial fitting on two-qubit ELT data. (a) Representative transformed PTM time series and overlaid fits for three parameters (e.g., $a_{z}^{(1)}$, $a_{zz}$, $D_{zzzz}$); slopes at the intercept define the recovered Lindblad parameters. 
    (b) Histogram of parameter differences (robust minus $\chi^2$), centered near zero with width comparable to the propagated $\chi^2$ uncertainties. 
    (c) Cumulative distribution functions of one–$\sigma$ uncertainties from $\chi^2$ fits and robust bounds computed via Eq.~\eqref{eq:RobustErrorBetter}.}
    \label{fig:L1Linffit}
\end{figure}

\paragraph{Guarantees and derivative bounds.}
Under the contamination assumption (at least half of samples within $\sigma$ of a true polynomial $q$), the robust fit satisfies
\begin{equation}
\max_{t \in [t_{\min}, t_{\max}]} \big| p_{\mathrm{tot}}(t) - q(t) \big| \;\le\; 3\sigma.
\end{equation}
França et\,al.~further derive a bound on the derivative at intercept (with high probability)
\begin{equation}\label{eq:RobustError}
    \big| p_{\mathrm{tot}}'(0) - q'(0) \big| \;\le\; \frac{3e\,\sigma}{t_{\min}}.
\end{equation}
While the bound in Eq.~\eqref{eq:RobustError} scales with $1/t_{\min}$ (potentially loose for very short times), one can instead derive the alternative bound of the form
\begin{equation}\label{eq:RobustErrorBetter}
\big| p_{\mathrm{tot}}'(0) - q'(0) \big| \;\le\; 3\sigma\, \frac{\bar{t}}{t_{\Delta}^2},
\end{equation}
which better mitigates the small–$t_{\min}$ sensitivity. Eq.~\eqref{eq:RobustErrorBetter} has not been shown in \cite{Franca2024Efficient}, but the proof technique is very close to being the same as for Eq.~\eqref{eq:RobustError}. The difference comes in how to get a bound on the derivative of a polynomial from just a bound on the polynomial itself. Concretely, it consists of using Theorem 2.20 of \cite{Chebyshev_polynomials} to get the derivative bound, analogous to what is done in \cite{Franca2025_time_dependent} instead of Markov brothers' inequality to get the same bound but possibly looser, which was originally done in \cite{Franca2024Efficient}. We will use Eq.~\eqref{eq:RobustErrorBetter} in quantifying uncertainty in Lindblad parameters extracted in the robust fitting technique, where we insert the upper bound for each time series that $\sigma \leq \max\limits_{i\in[T]}{\sigma_{i}}$. 

In this work, we adopt uniformly spaced idling times and ordinary $\chi^2$ regression (linear fits in the short-time regime), primarily for transparency and experimental reproducibility. 
To validate this choice, we applied both the robust scheme and $\chi^2$ fits to two-qubit ELT datasets and compared the recovered slopes (derivatives at the intercept) for multiple transformed PTMs. 
The results are summarized in Fig.~\ref{fig:L1Linffit}.

Empirically, the robust and $\chi^2$ estimates agree within uncertainties: The difference histogram in Fig.~\ref{fig:L1Linffit}(b) is centered near zero with a dispersion comparable to the $\chi^2$ error bars, and the CDFs of uncertainties $\chi^2$ greatly outperforms the error bound from Eq.~\eqref{eq:RobustErrorBetter}, as can be seen in Fig.~\ref{fig:L1Linffit}(c). 
We therefore use $\chi^2$ fits throughout the main text; the robust method provides a compatible alternative with explicit worst-case guarantees when contamination or heavy tails are of concern but which explicitly does not seem the be the case for our data. 


\twocolumngrid
\bibliography{references}

\end{document}